\newif\iffigframe
\newcommand{\figbox}[3]{\hbox to#1\bgroup
  \dimen0=1bp \dimen1=#1\relax
  \def\a##1 ##2 ##3 ##4 ##5\\{\if!##1!\a##2 ##3 ##4 ##5 .\\\else
    \dimen3=##3\dimen0 \advance\dimen3 -##1\dimen0
    \dimen4=##4\dimen0 \advance\dimen4 -##2\dimen0
    \dimen5=\dimen4 \divide\dimen5 \dimen3
    \dimen2=\dimen1 \multiply\dimen2 \dimen5
    \multiply\dimen5 \dimen3 \advance\dimen4 -\dimen5
    \dimen5=\dimen1
    \loop \advance\dimen4 \dimen4 \divide\dimen5 2
    \ifnum\dimen5>0 \ifnum\dimen4<\dimen3 \else
      \advance\dimen4 -\dimen3 \advance\dimen2 \dimen5 \fi
    \repeat
    \dimen5=10\dimen1 \divide\dimen5 \dimen0
    \includegraphics{#3.eps}%
    \iffigframe \vrule\hss \else \hfil \fi
    \vbox to\dimen2\bgroup
      \iffigframe \hrule width\dimen1\vss \hrule \else \vfil \fi
      \egroup
    \iffigframe \vrule\hss \fi
    \egroup\fi}%
  \a#2 . . . .\\}
 
\documentstyle[11pt,a4]{article}
 
\newlength{\dinwidth}
\newlength{\dinmargin}
\def\eqalign#1{
\null \,\vcenter {\openup \jot \ialign {\strut \hfil $\displaystyle {
##}$&$\displaystyle {{}##}$\hfil \crcr #1\crcr }}\,}
 
\def\abs#1{\vert #1\vert}
\def\begeq#1{\begin{equation}\label{#1}}
\def\endeq{\end{equation}}
\newcounter{subequation}[equation]
 
\makeatletter
\expandafter\let\expandafter\reset@font\csname reset@font\endcsname
\newenvironment{subeqnarray}
  {\arraycolsep1pt
    \def\@eqnnum\stepcounter##1{\stepcounter{subequation}{\reset@font\rm
      (\theequation\alph{subequation})}}\eqnarray}%
  {\endeqnarray\stepcounter{equation}}
\makeatother
 
\begin{document}
\vspace{1cm}
\begin{center}
\LARGE\bf
Static spherically symmetric monopole solutions in the presence of a
dilaton field
\vspace{10mm}
 
\large Peter Forg\'acs {\normalsize and} J\'ozsef Gy\"ur\"usi
 
\vspace{3mm}
\small\sl
Research Institute for Particle and Nuclear Physics\\
P.O.Box 49\\
H-1525 Budapest 114, Hungary
\vspace{20mm}
\end{center}
 
\begingroup\addtolength{\leftskip}{1cm}\addtolength{\rightskip}{1cm}
 
\subsection*{Abstract}
A numerical study of static, spherically symmetric
monopole solutions of a spontaneously broken SU(2) gauge theory
coupled to a dilaton field is presented.
Regular solutions seem to exist only up a maximal value of
the dilaton coupling.
In addition to the generalization of the 't~Hooft-Polyakov
monopole a discrete family of regular solutions is found,
corresponding to radial excitations absent in the theory
without dilaton.
 
\endgroup
\newpage
 
\noindent
The aim of this paper is to present a detailed numerical
study of classical solutions of an
SU(2) Yang-Mills-Higgs (YMH) theory coupled to a
 (massless) dilaton (DYMH),
when the Higgs field is in the adjoint representation.
Dilaton fields appear naturally in low energy effective
field theories derived from superstring models \cite{Wi, BFQ, GSW}.
As previous studies have already shown \cite{LavI, BizI} the
inclusion of a dilaton in a pure Yang-Mills (YM) theory
has drastic consequences already at the classical level.
In particular the dilaton Yang-Mills (DYM)
theory possesses finite energy `particle-like'
solutions which are absent in the pure YM case.
The same phenomenon happens in the
Einstein-Yang-Mills system where non-singular finite
energy solutions have been discovered some time ago \cite{BM}.
 
The above YMH model without a dilaton field
is known to possess nonsingular, finite energy
solutions, describing magnetic monopoles \cite{THP}.
In the present work we study static, spherically symmetric
solutions of the DYMH theory.
Our results show a striking similarity to those obtained
in the corresponding YMH theory coupled to gravity (EYMH)
\cite{BFM}.
In the DYMH theory
there is a finite energy abelian solution for all values of
the dilaton coupling {$\alpha$}.
Based on numerical investigations there is a strong
indication for the existence of a finite energy
{\sl nonabelian} monopole up to only
a maximal value of $\alpha$, $\alpha_{\rm m}$. 
The nonabelian monopole merges with the abelian
one at a critical value of $\alpha$, $\alpha_{\rm c}$.
In limit when the dilaton decouples the nonabelian
solution joins smoothly 
the 't Hooft-Polyakov monopole. If the Higgs self
coupling, $\beta$, is
in the interval [0,0.6] $\alpha_{\rm m}(\beta)$ and
$\alpha_{\rm c}(\beta)$
are different and with close analogy to the EYMH theory there are two
different solutions if  $\alpha_{\rm c}<\alpha<\alpha_{\rm m}$.
 
In addition to the `fundamental' monopole there is 
good numerical evidence that 
a countable family
of globally regular solutions exits 
for $0<\alpha<\sqrt3/2$ independently of
$\beta$.
As their energy is
higher they can be interpreted as excitations of the
fundamental monopole.
In the limit when the dilaton filed decouples
their energy diverges.
There is another limit when the DYMH theory reduces to the
DYM model and
then these excitations tend to the solutions discovered
in refs.~\cite{LavI,BizI}.

The action of our model is given by
 \begeq{sdymh}
   S=\int d^4x\left\{{1\over2}\partial_\mu\varphi\,\partial^\mu\varphi-
   e^{2\kappa \,\varphi}{1\over4g^2}F_{\mu\nu}^aF^{a\mu\nu}+
   {1\over 2}(D_\mu\Phi)^a (D^\mu\Phi)^a-e^{-2\kappa \,\varphi}\,V(\Phi)
   \right\}\,,
 \endeq
 where $a=1$,$2$,$3$, and the Higgs potential is
  \begeq{higgspot}
     V(\Phi)={\lambda\over8}(\Phi^a\Phi^a-v^2)^2\,.
  \endeq
 The couplings of the scalar field $\varphi$ has been chosen so that the
 action (\ref{sdymh}) possesses the following dilatational symmetry
 \begeq{dymhdil}
   x^\mu\to e^{\kappa  c}x^\mu\;,\quad A_\mu\to e^{-\kappa  c}A_\mu\;,
   \quad \varphi\to\varphi+c\;,\quad S\to e^{2\kappa  c}S\,,
 \endeq
 which can be used to fix the value of $\varphi$ arbitrarily at any point.
 With the
 \begeq{sscale}\eqalign{
  \Phi\to v\Phi,\quad x\to{1\over gv}x,&\quad\varphi\to v\varphi,
  \quad S\to{v\over g}S,\cr
  \lambda\to{g^2}\beta^2,&\quad\kappa \to{1\over v}\alpha}
 \endeq
 rescaling of variables the only remaining dimensionless parameters are
 \begeq{alphabeta}
   \alpha={\kappa M_W\over g}\;,\quad \beta={M_H\over M_W}\;,
 \endeq
 where $M_W=vg$ and $M_H=v\sqrt{\lambda}$ are the
 characteristic masses of the theory.
 Without further restrictions one can assume $\alpha\geq0$.
 Let us mention here that the action of the DYMH theory (\ref{sdymh})
 can be obtained from the Einstein-Yang-Mills-Higgs
 action by restricting the
 metric to the special conformstatic form
 $ds^2=e^{2\phi}dt^2-e^{-2\phi}dx^idx^i$.
 
 We use the minimal static spherically symmetric ansatz with zero
 `electric' field
   \begin{subeqnarray}\label{minanz}
     A^a_0\equiv 0,\quad
     A^a_i&=&\epsilon_{aij}{x^j\over r^2}(W(r)-1)\;,\\
     \Phi^a&=&H(r){x^a\over r}\;,\\
     \varphi&=&\varphi(r)\;.
   \end{subeqnarray}
  After the rescaling (\ref{sscale}) the action (\ref{sdymh}) reduces
 to
 \begeq{sdymhanz}
E=\int dr\left\{{1\over2}r^2{\varphi'}^2+e^{2\alpha\varphi}
\Bigl[
    {W'}^2+{(W^2-1)^2\over2r^2}\Bigr]+W^2H^2+
{r^2{H'}^2\over2}+
    {\beta^2\over8}r^2e^{-2\alpha\varphi}(H^2-1)^2\right\}
 \endeq
 using the ansatz (\ref{minanz}).
 Varying (\ref{sdymhanz}) we obtain the equations of motion:
 \begin{subeqnarray}\label{dymhanzeq}
    \Bigl(r^2\varphi'\Bigr)'&=&2\alpha e^{2\alpha\varphi}\Bigl({W'}^2+
       {(W^2-1)^2\over2r^2}\Bigr)-{\alpha\beta^2\over4}r^2e^{-2\alpha\varphi}
       (H^2-1)^2\;,\\
   r^2W''&=&W(W^2-1+r^2e^{-2\alpha\varphi}H^2)-2\alpha r^2\varphi'W'\;,\\
   (r^2H')'&=&H\Bigl(2W^2+{\beta^2\over2}r^2e^{-2\alpha\varphi}(H^2-1)\Bigr)
   \;.
 \end{subeqnarray}
 Solutions regular at the origin must satisfy the following boundary
 conditions (b.c.):
 \begeq{dymhsor}\eqalign{
   H=ar+O(r^2),\quad W&=1-br^2+O(r^3),\quad \cr
   \varphi=\varphi_0+\alpha\Bigl(2b^2e^{2\alpha\varphi_0}&-
   {\beta^2\over24}e^{-2\alpha\varphi_0}\Bigr)r^2+O(r^3)\;,}
 \endeq
 i.e.~there is a three parameter ($a$, $b$, $\varphi_0$)
 family of regular solutions at $r=0$. The local existence can be proved
 following the procedure discussed in \cite{BFMII}
 
 For $r\to\infty$ the corresponding `regular' b.c. are
 \begin{subeqnarray}\label{dymhinfreg}
  \varphi&=&\varphi_\infty-{d\over r}+O({1\over r^2})\;,\\
  W&=&B e^{-\mu\rho}\left(1+O({1\over r})\right)\;,\\
  H&=&1-{C\over r}e^{-\mu\beta\rho}\left(1+O({1\over r})
       \right)\quad{\rm for}\;\;\beta<2\\
  H&=&1-{2B^2\over {\mu^2(\beta^2-4)r^2}}e^{-2\mu\rho}
       \left(1+O({1\over r})\right)\quad{\rm for}\;\;\beta>2\;,
 \end{subeqnarray}
 parametrized by ($\varphi_\infty$, $d$, $B$, $C$), where
 $\mu=e^{-\alpha\varphi_\infty}$, $\rho=r+\alpha d \ln r$.
 For $\beta>2$ we cannot fully parametrize the stable
 manifold at $r=\infty$.
 Exploiting the virial theorem and using Eqs.~(\ref{dymhanzeq}),
 (\ref{dymhsor}) and (\ref{dymhinfreg}) one obtains for the energy
 (\ref{sdymhanz})
 \begeq{dymhvire}
    E={1\over\alpha}\int\limits_0^\infty dr\,(r^2\varphi')'=
      {1\over\alpha}\left.(r^2\varphi')\right|_0^\infty.
 \endeq
 
 We discuss first an exact two parameter family of solutions of
 Eqs.~(\ref{dymhanzeq}) which is going to play
 an important role in our further analysis,
 since it descibes the asymptotic behaviour of nonabelian solutions.
 Consider the singular abelian monopole, $W\equiv0$, $H\equiv1$ then
 the general solution of (\ref{dymhanzeq}) takes the form
 \begeq{dymhgenrn}
  \varphi_{\rm a}=-{1\over\alpha}\ln\abs{{1\over A}\sinh\Bigl(A({1\over M}+
       {\alpha\over r})\Bigr)}
 \endeq
 with  $A^2,M\in{\bf R}$.
 For $M>0$ and $A$ real it is regular for all $r>0$.
 Its energy, $E$ Eq.~(\ref{sdymhanz}), is infinite unless $A=0$. For $A=0$
 the total energy of the solution (\ref{dymhgenrn}) is finite, $E=M/\alpha$.
 When $M<0$ and $A$ is real the solution is singular at $r=-\alpha M$ and
 the energy diverges.
 If $A$ is imaginary for any value of $M$ the solution becomes singular at
 some $r>0$.
 
 Using the dilatational symmetry (\ref{dymhdil}) from now on we set
 $\varphi_\infty=0$. Note that then the energy scales as
 $E\to \exp(-\alpha\varphi_\infty)E$. The finite energy abelian solution
 for $\varphi_\infty=0$ reads
$$\varphi=-{1\over\alpha}\ln(1+{\alpha\over r})\,,$$
  and its energy is  simply $E=1/\alpha$.
 It satisfies the following first order differential Eq.
 \begeq{dymhbogeq}
   r^2\varphi'=\pm e^{\alpha\varphi}\,,
 \endeq
 which is derivable from a Bogomolny type bound \cite{BizI} in the
 background $W\equiv0$, $H\equiv1$.
 
 The asymptotic behaviour of finite energy nonabelian solutions is described
 by (\ref{dymhgenrn}). With the help of Eq.~(\ref{dymhvire})
 we arrive at the following useful formula for the total energy
 of globally regular nonabelian solutions:
 \begeq{dymhaszenergscal}
     E={1\over\alpha}\cosh(A/M)\;,
 \endeq
 where $A$, $M$ are parameters of $\varphi_{\rm a}$ describing their
 asymptotic behaviour.
 For globally regular solutions of (\ref{dymhanzeq}) satisfying
 the b.c.~(\ref{dymhsor}), (\ref{dymhinfreg}) the energy (\ref{sdymhanz})
 is a monotonically increasing function of the Higgs coupling $\beta$ and
 monotonically decreases with the dilaton coupling $\alpha$.
 (We remark here that the condition $\varphi_\infty=0$ is
 essential to prove these facts.)
 
 We have numerically integrated a suitably desingularized version of
 Eqs.~(\ref{dymhanzeq}) from $r=0$ using the b.c.~(\ref{dymhsor}).
 We set $\varphi_0$ to zero in order to have to vary only two parameters.
 By adjusting the parameters $a$, $b$ we suppressed the divergent modes
 in the asymptotic region according to (\ref{dymhaszmod}). The resulting
 solutions can then be transformed to satisfy $\varphi_\infty=0$ by the
 dilatational symmetry (\ref{dymhdil}).
 
\begin{figure}
 \hbox to\hsize{\hss
  \figbox{0.5\hsize}{70 72 818 556}{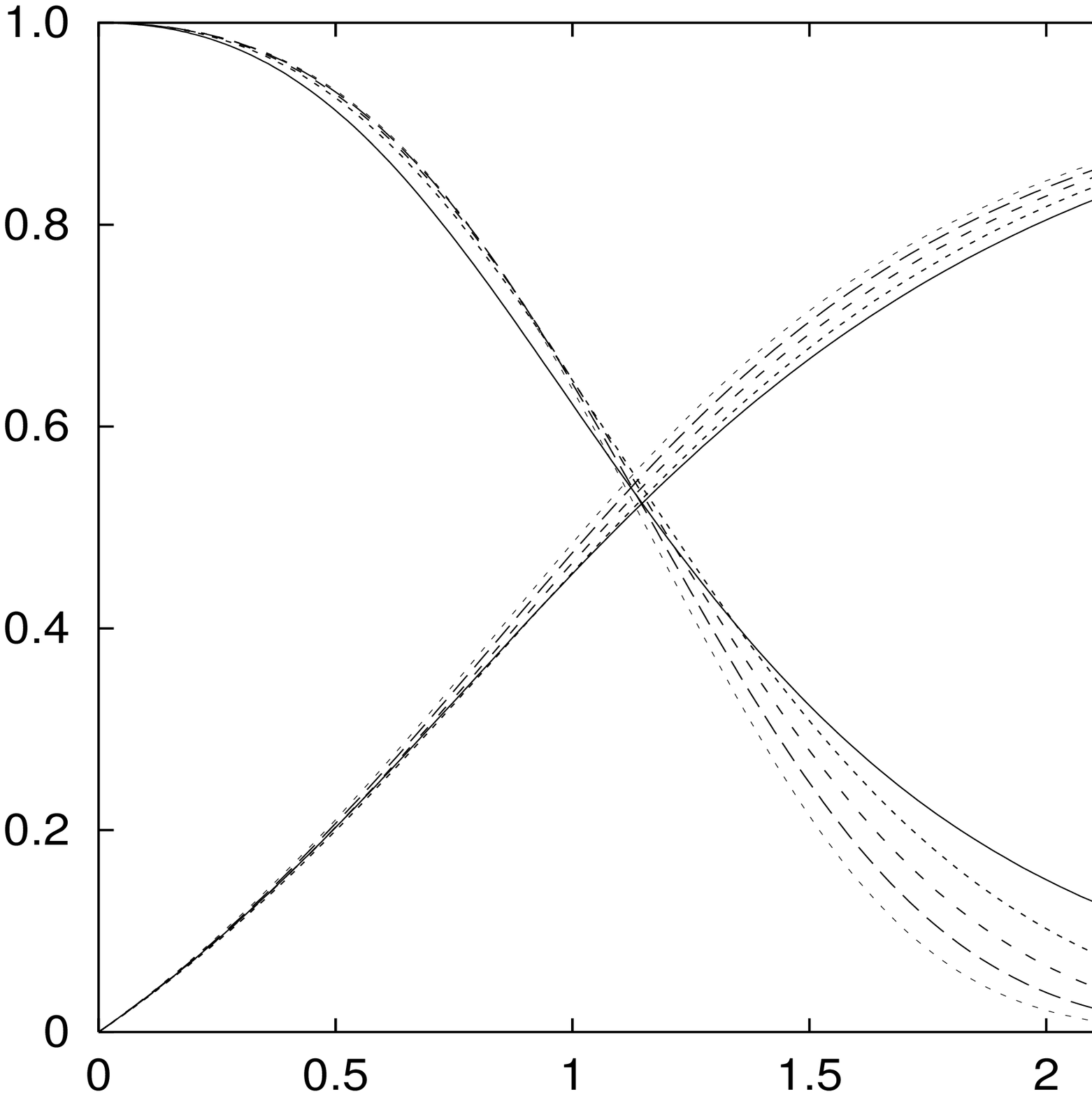}\hss
  \figbox{0.5\hsize}{70 72 818 556}{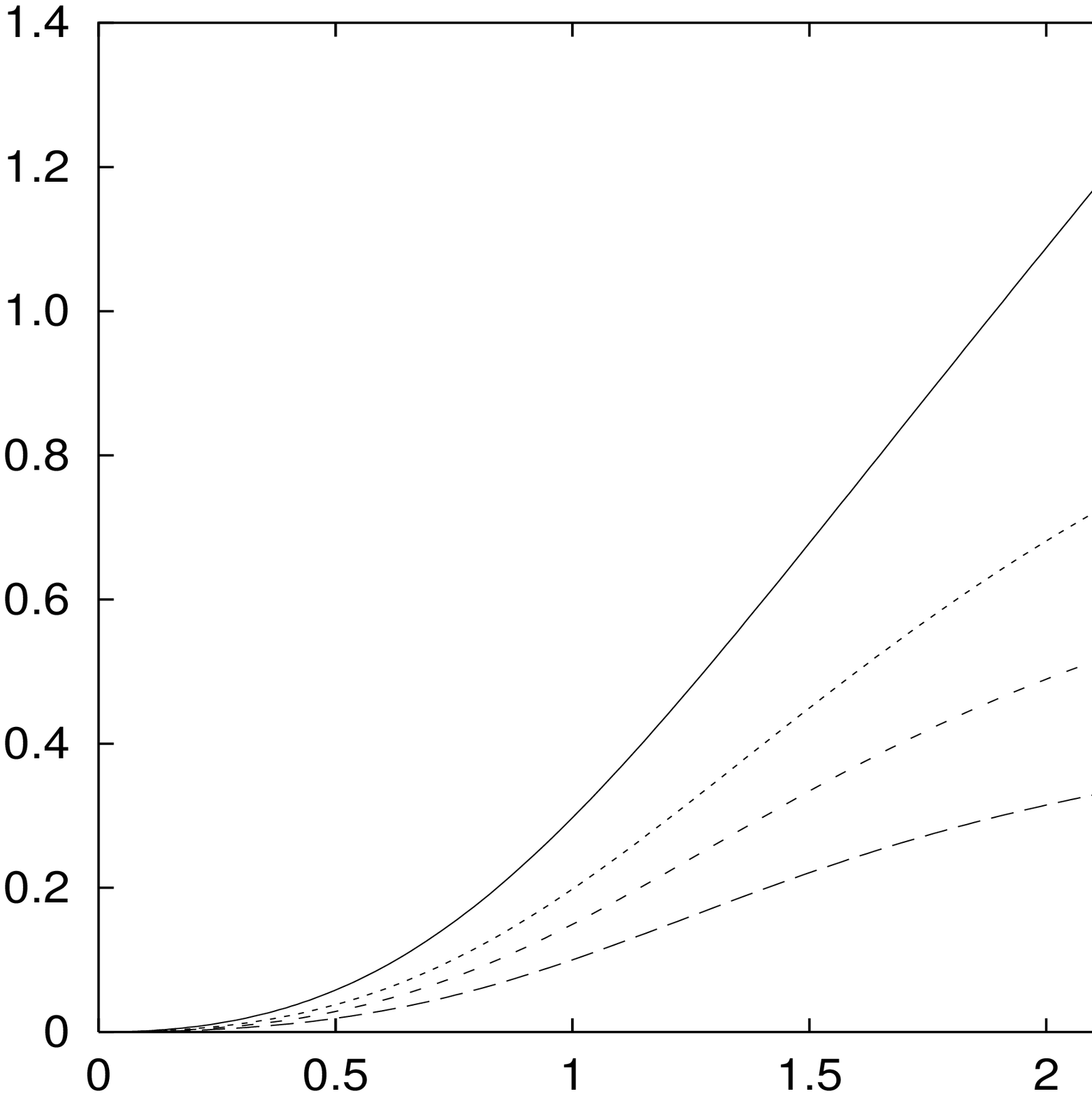}\hss }
\begin{picture}(0,0)(0,0)
\put(205,0){\makebox(0,0){\footnotesize$\ln(1+r)$}}
\put(435,0){\makebox(0,0){\footnotesize$\ln(1+r)$}}
\put(25,170){\makebox(1,1){\small\footnotesize$W$, \footnotesize$H$}}
\put(250,170){\makebox(1,1){\small$\varphi$}}
\end{picture}
 \caption[0b0c]{\label{0b0c}
 The $\alpha$ dependence of the fundamental monopole solution for
 $\beta=0$}
\end{figure}

 We found that there exist a fundamental nonabelian monopole solution
 for all $\beta$ below a maximal value of $\alpha$, $\alpha_{\rm m}$
 which smoothly joins 
 the 't~Hooft-Polyakov monopole in the $\alpha\to 0$ limit.
 Some solution curves for $\beta=0$ are shown in Fig.~\ref{0b0c}.
 For the parameters see Table~{\ref{T1}}.
\begin{table}[htb]
\begin{minipage}[t]{10.3cm}
\caption{}
\vspace{.2truecm}
\label{T1}
{\small
\begin{tabular}{|l|c|c|l|c|}
\hline
\multicolumn{5}{|c|}{Parameters for the fundamental monopole
solutions in Fig.~\ref{0b0c}}\\ \hline
\multicolumn{1}{|c|}{$\alpha$}&
\multicolumn{1}{|c|}{$a$}&
\multicolumn{1}{|c|}{$b$}&
\multicolumn{1}{|c|}{$\phi_\infty$}&
\multicolumn{1}{|c|}{$\alpha\cdot E$}\\ \hline\hline
 0.0     & 0.33333333 & 0.16666666 & 0  & 0{\phantom .}~~~~~~~~~ \\
 0.1     & 0.33324422 & 0.16674419 & 0.050094 & 0.099853   \\
 0.2     & 0.33297626 & 0.16697893 & 0.100758  & 0.198827    \\
 0.5     & 0.33107418 & 0.16872021 & 0.262673  & 0.481732    \\
 0.7     & 0.32884174 & 0.17095758 & 0.387902  & 0.650049    \\
 0.8     & 0.32741363 & 0.17252229 & 0.460048  & 0.725582    \\
 1.0     & 0.32388447 & 0.17699856 & 0.638684  & 0.855180    \\
 1.1     & 0.32176986 & 0.18029088 & 0.758343  & 0.907479   \\
 1.2     & 0.31944319 & 0.18485325 & 0.919110  & 0.950109    \\
 1.3     & 0.31705890 & 0.19195967 & 1.172960   & 0.981707    \\
 1.4     & 0.31657705 & 0.20987214 & 1.950448   & 0.999523    \\
 1.4088  & 0.31898696 & 0.21924215 & 2.582925   & 1.000092     \\
 1.4     & 0.32380520 & 0.22987901 & 5.289654   & 1.000000     \\
 1.39939 & 0.32402298 & 0.23025275 & 8.461582   & 1.000000     \\
\hline
\end{tabular}
}
\end{minipage}
\hskip .5cm
\begin{minipage}[t]{5cm}
\caption{}
\vspace{.2truecm}
\label{T4}
{\small
\begin{tabular}{|l|c|r@{.}l|}
\hline
\multicolumn{1}{|c|}{$\beta ^2$}&
\multicolumn{1}{|c|}{$\alpha_{\rm c}$}&
\multicolumn{2}{|c|}{$\alpha_{\rm m}$} \\ \hline\hline
 
 0        & 1.39938 &       1&4088 \\
 0.02     & 1.37874 &       1&3803 \\
 0.03     & 1.36950 &       1&3702 \\
 0.04     & 1.36088 &       1&3612 \\
 0.043    & 1.35839 &       1&3586 \\
 0.05     & 1.35279 &       1&3529 \\
 0.06     & $\alpha_{\rm m}$ & 1&3452 \\
 0.07     & $\alpha_{\rm m}$ & 1&338 \\
 0.08     & $\alpha_{\rm m}$ & 1&33 \\
 0.1      & $\alpha_{\rm m}$ & 1&32 \\
 0.15     & $\alpha_{\rm m}$ & 1&29 \\
 0.2      & $\alpha_{\rm m}$ & 1&27 \\
 1        & $\alpha_{\rm m}$ & 1&11 \\
 4        & $\alpha_{\rm m}$ & 0&98\\
 $\infty$ & $\alpha_{\rm m}$ & 0&89\\
\hline
\end{tabular}
}
\end{minipage}
\end{table}
 
There is a critical $\alpha$ value, $\alpha_{\rm c}$,
where the dilaton field
becomes logarithmically divergent while $W$ and $H$ tend to
some nontrivial
functions as can be seen on Fig.~\ref{0b0c}. After shifting
 $\varphi_\infty$
to 0 the fundamental monopole tends to the finite energy abelian one as
$\alpha\to\alpha_{\rm c}$.
For $\beta^2\leq0.06$ $\alpha_{\rm c}$
is seen to differ from $\alpha_{\rm m}$
(see Table \ref{T4}).
It means that for a given $\alpha$ in the interval
[$\alpha_c,\alpha_{\rm m}$] there are two different nonabelian solutions
with different energies. The function
$\alpha_{\rm m}(\beta)$  decreases with increasing $\beta$ from
$\alpha_{\rm m}(0)\approx1.4088$ to $\alpha_{\rm m}(\infty)\approx0.89$.
 
\begin{figure}
 \hbox to\hsize{\hss
  \figbox{0.5\hsize}{70 72 818 556}{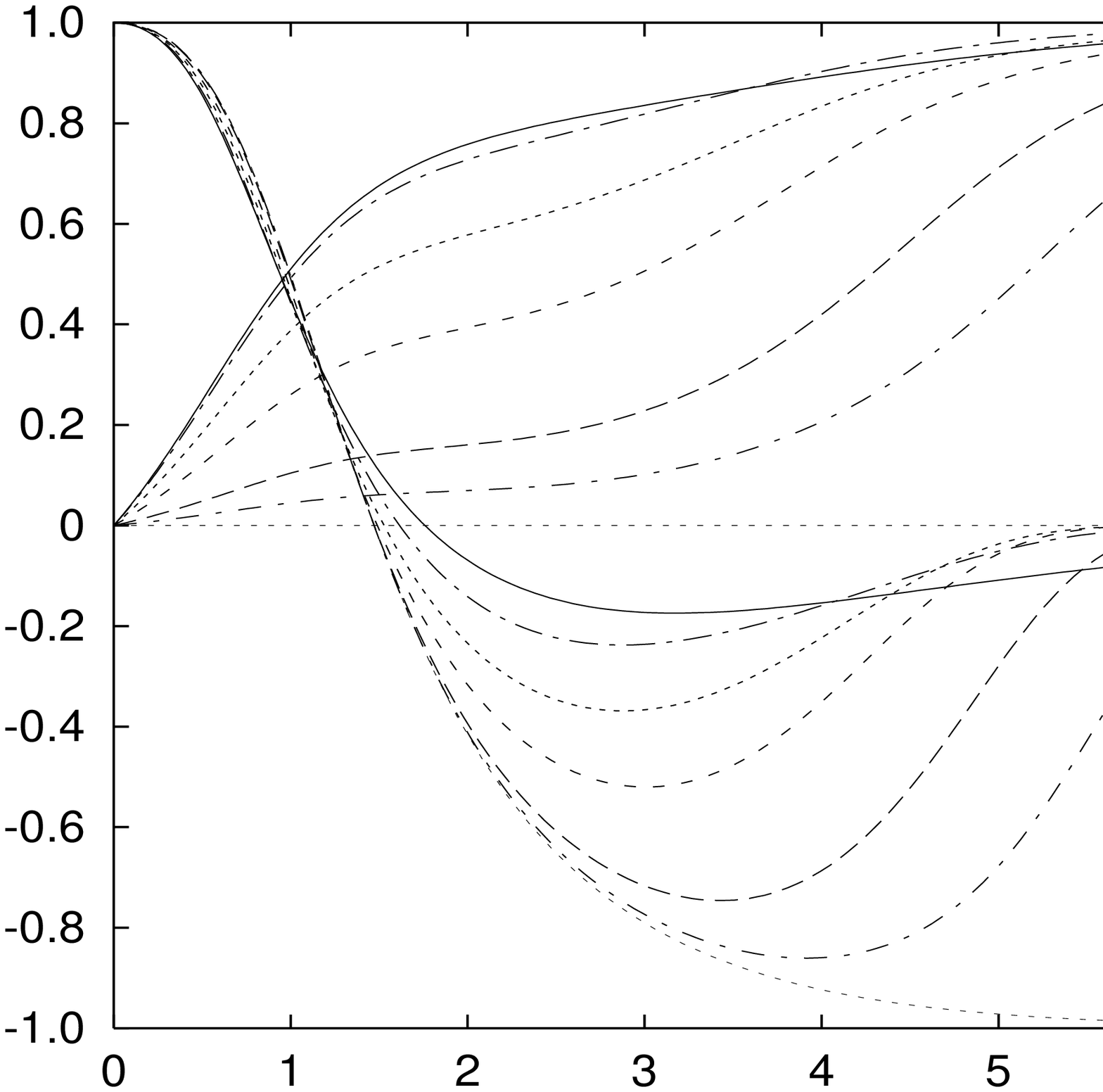}\hss
  \figbox{0.5\hsize}{70 72 818 556}{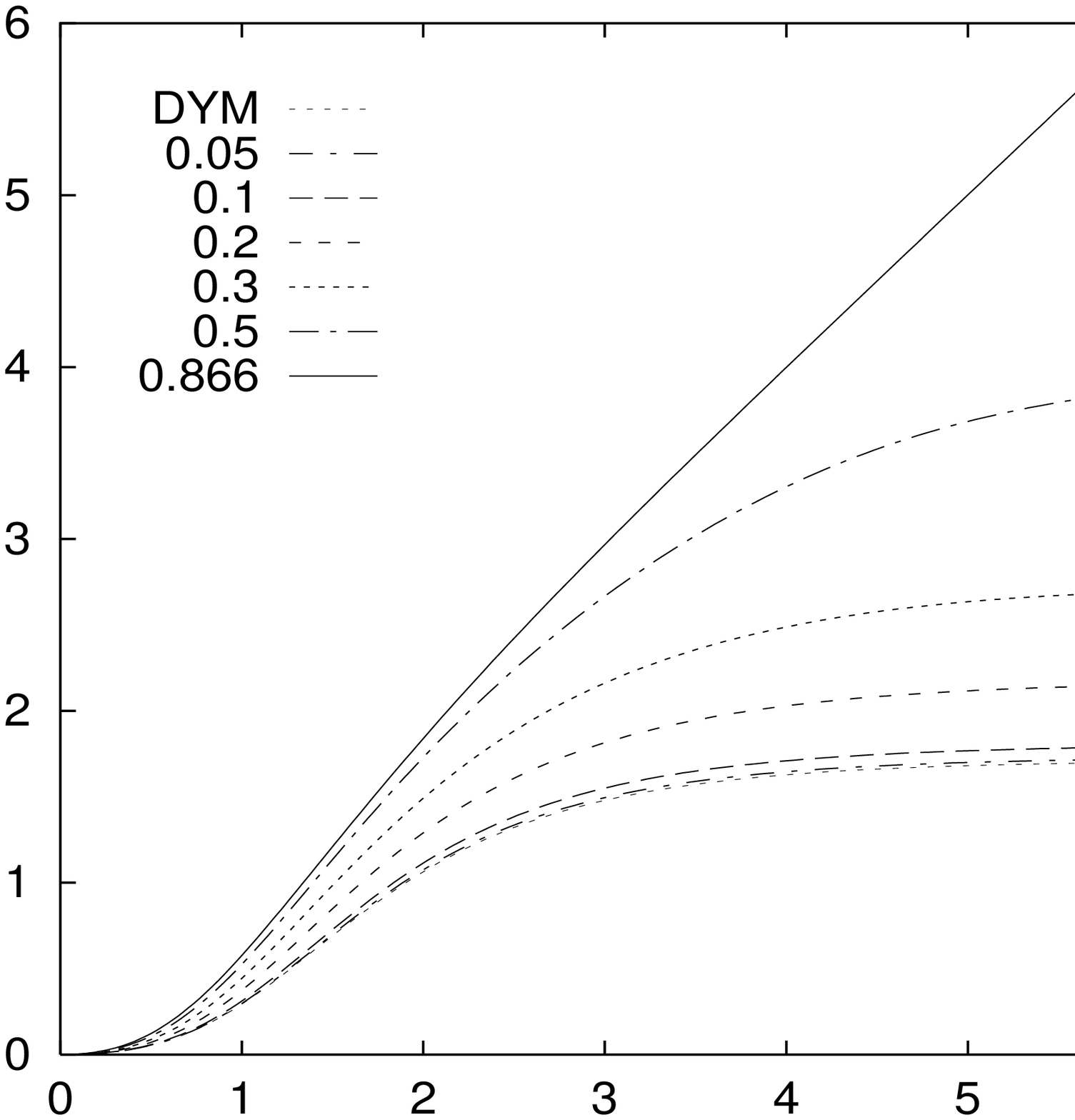}\hss }
\begin{picture}(0,0)(0,0)
\put(205,0){\makebox(0,0){\footnotesize$\ln(1+r/\alpha)$}}
\put(435,0){\makebox(0,0){\footnotesize$\ln(1+r/\alpha)$}}
\put(25,170){\makebox(0,0){\small\footnotesize$W$, \footnotesize$H$}}
\put(250,170){\makebox(0,0){\small$\alpha\varphi$}}
\end{picture}
  \caption[0b1c]{\label{0b1c}
 The $\alpha$ dependence of the first excited monopole solution for
 $\beta=0$}
\end{figure}

\begin{figure}
 \hbox to\hsize{\hss
  \figbox{0.5\hsize}{70 72 818 556}{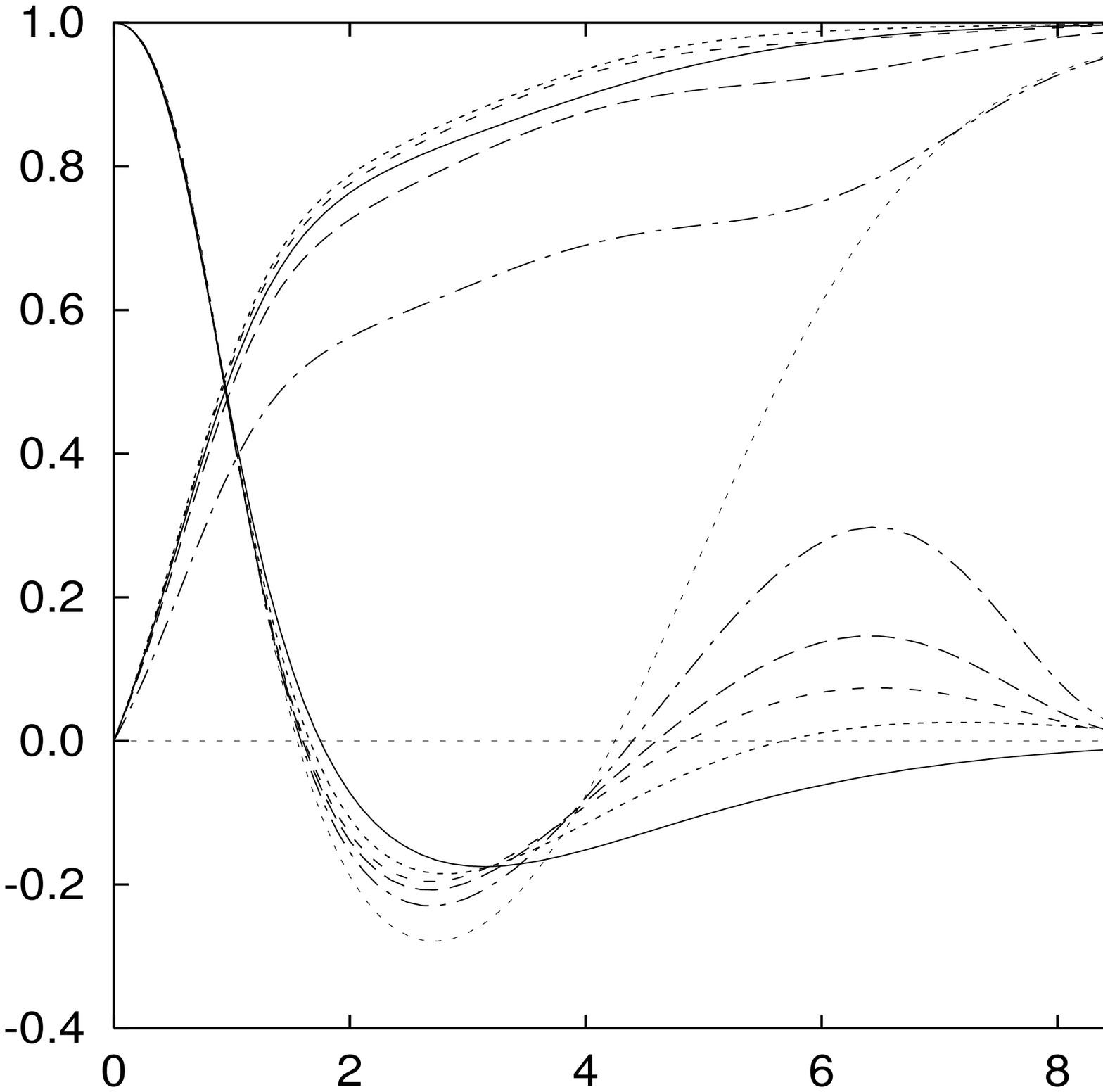}\hss
  \figbox{0.5\hsize}{70 72 818 556}{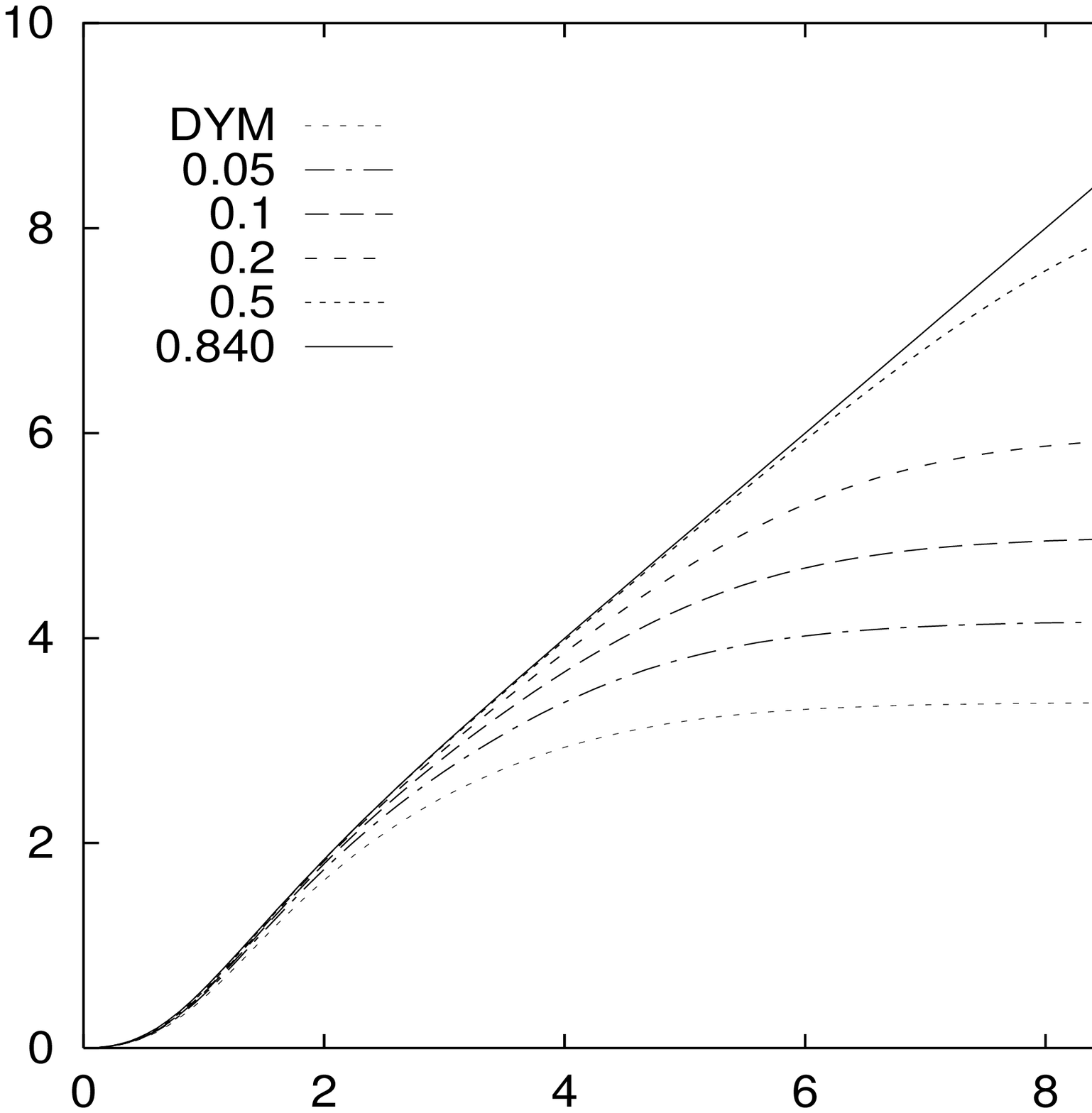}\hss }
\begin{picture}(0,0)(0,0)
\put(205,0){\makebox(0,0){\footnotesize$\ln(1+r/\alpha)$}}
\put(435,0){\makebox(0,0){\footnotesize$\ln(1+r/\alpha)$}}
\put(25,170){\makebox(0,0){\small\footnotesize$W$, \footnotesize$H$}}
\put(250,170){\makebox(0,0){\small$\alpha\varphi$}}
\end{picture}
 \caption[0b2c]{\label{0b2c}
 The $\alpha$ dependence of the second excited monopole solution for
 $\beta=0$}
\end{figure}

There also seems to exist a countable family of globally regular monopole
solutions indexed by the number zeros of $W(r)$ for all $\beta$
and $0<\alpha<\sqrt3/2$. They can be interpreted as radial excitations of
the fundamental monopole. In the $\alpha\to 0$ limit after a suitable
rescaling they can be identified with the previously discovered DYM
solutions \cite{LavI},\cite{BizI}.
We illustrate some of these excited
solutions with one and two zeros for $\alpha$ various values 
on Figs.~\ref{0b1c} and \ref{0b2c}, while the corresponding
parameters are listed in Table~\ref{T2} and Table~\ref{T3}.
\begin{table}[htb]
\centering
\caption{}
\vspace{.2truecm}
\label{T2}
\begin{tabular}{|l|l|l|r@{.}l|c|}
\hline
\multicolumn{6}{|c|}{Parameters for the first excited monopole
solution in Fig.~\ref{0b1c}}\\ \hline
\multicolumn{1}{|c|}{$\alpha$}&
\multicolumn{1}{|c|}{$a$}&
\multicolumn{1}{|c|}{$\alpha^2b$}&
\multicolumn{2}{|c|}{$\alpha\varphi_\infty$}&
\multicolumn{1}{|c|}{$\alpha\cdot E$} \\ \hline\hline
 
 {\small DYM } & ~~~~~~---& 0.26083015 & 1&711412 & 0.8038078 \\
 0.05 & 0.66718265 & 0.26286356 & 1&731264 & 0.8507008 \\
 0.1  & 0.77749002 & 0.27004539 & 1&803886 & 0.8904364 \\
 0.2  & 0.99147112 & 0.30099818 & 2&170909 & 0.9448188 \\
 0.3  & 1.00351436 & 0.33435729 & 2&730953 & 0.9720677 \\
 0.5  & 0.77902186 & 0.37296818 & 3&994232 & 0.9932192 \\
 0.7  & 0.59081576 & 0.39281185 & 5&919563 & 0.9992733 \\
 0.8  & 0.51698205 & 0.39916099 & 8&207917 & 0.9999547 \\
 0.866& 0.47164657 & 0.40215328 & 26&31413 & 1.0000000  \\
\hline
\end{tabular}
\end{table}
\begin{table}[htb]
\centering
\caption{}
\vspace{.2truecm}
\label{T3}
\begin{tabular}{|l|l|l|l|c|}
\hline
\multicolumn{5}{|c|}{Parameters for the second excited monopole
solution in Fig.~\ref{0b2c}}\\ \hline
\multicolumn{1}{|c|}{$\alpha$}&
\multicolumn{1}{|c|}{$a$}&
\multicolumn{1}{|c|}{$\alpha^2b$}&
\multicolumn{1}{|c|}{$\alpha\varphi_\infty$}&
\multicolumn{1}{|c|}{$\alpha\cdot E$} \\ \hline\hline
 
 {\small DYM}  & ~~~~~~~---    & 0.35351801 & 3.373903 & 0.96559852 \\
 0.05 & 6.0251373  & 0.36759603 & 4.169777 & 0.99251583 \\
 0.1  & 3.9102367  & 0.37450939 & 4.996143 & 0.99673418 \\
 0.2  & 2.0946637  & 0.37885368 & 5.999793 & 0.99882187 \\
 0.3  & 1.4148625  & 0.38161925 & 6.815222 & 0.99949621 \\
 0.5  & 0.85071037 & 0.38762800 & 8.666195 & 0.99993064 \\
 0.7  & 0.60036405 & 0.39522735 & 12.25580 & 0.99999862 \\
 0.8  & 0.51818563 & 0.39944391 & 17.77525 & 1.00000000 \\
 0.84 & 0.48948129 & 0.40111050 & 25.77133 & 1.00000000  \\
\hline
\end{tabular}
\end{table}
 
 When $\alpha\to \sqrt3/2$ for the excited solutions the dilaton
 diverges logarithmically again while the zeros of $W$ go rapidly to
 infinity, except for the innermost one.
 For $\sqrt3/2<\alpha<1$ there is a surviving solution with
 a single zero of $W$ and with
 divergent $\varphi$ existing up to $\alpha=1$, where $W$ 
 develops an extra divergent mode, so for $\alpha>1$ this
 solution is not expected to exist.
 There also exists
 another type of limiting solution when the number of zeros of $W$ goes to
 infinity for all $0<\alpha<\sqrt3/2$ and the dilaton field diverges
 logarithmically.
 
\begin{figure}
 \hbox to\hsize{\hss
  \figbox{0.5\hsize}{70 72 818 556}{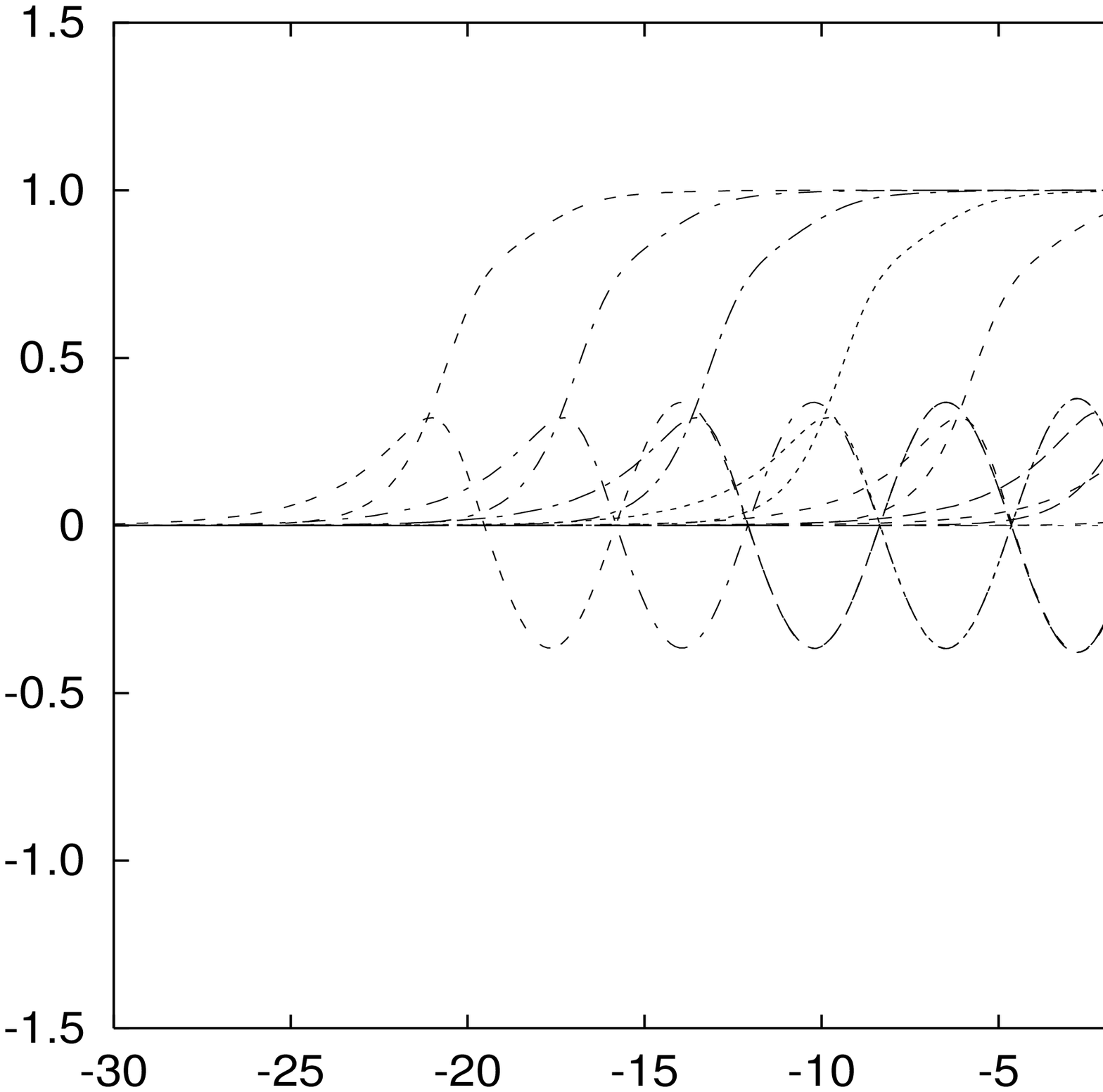}\hss
  \figbox{0.5\hsize}{70 72 818 556}{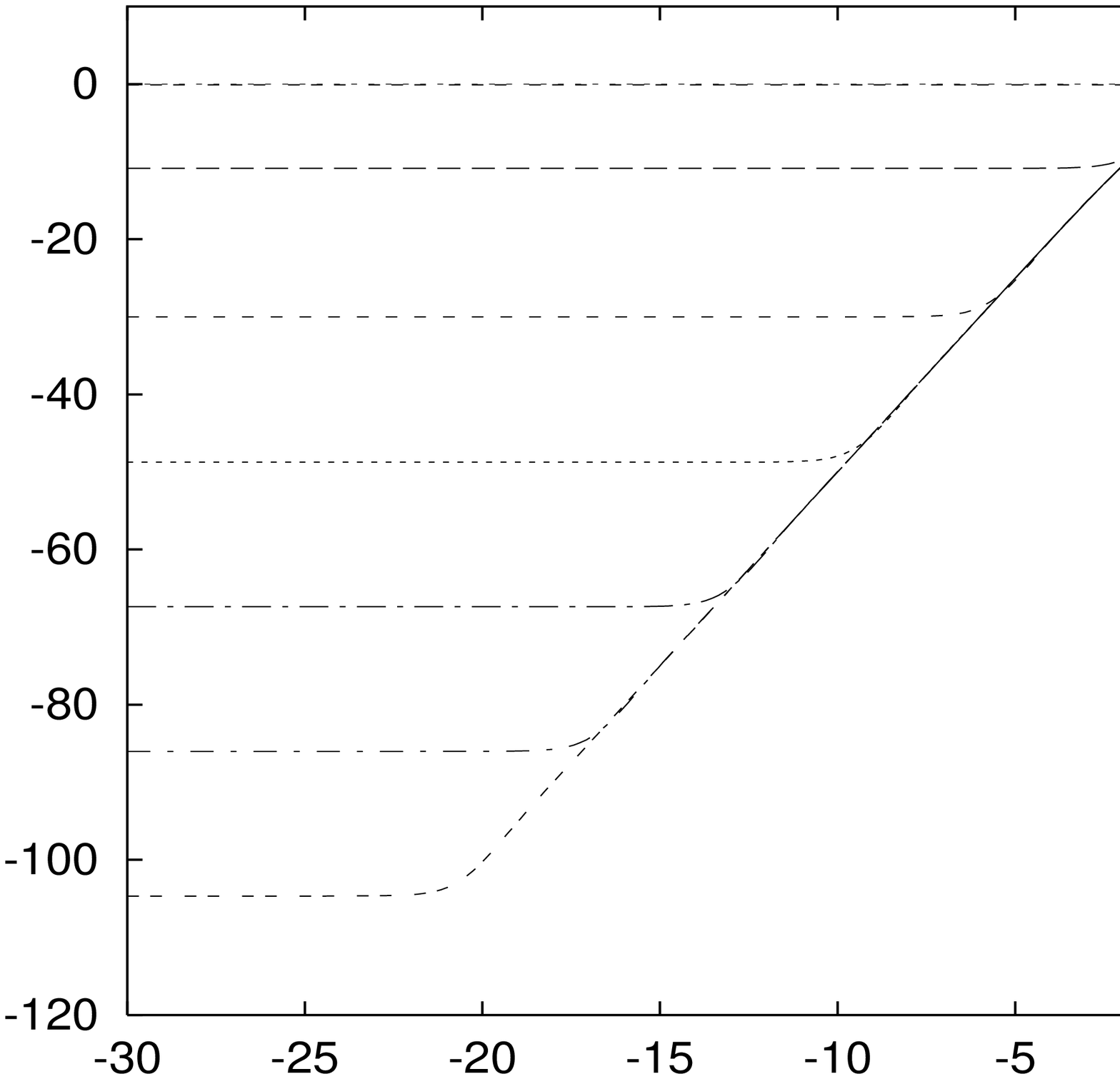}\hss }
\begin{picture}(0,0)(0,0)
\put(200,0){\makebox(0,0){\small$\ln(r/\alpha)\!-\!\alpha\varphi_\infty$}}
\put(425,0){\makebox(0,0){\small$\ln(r/\alpha)\!-\!\alpha\varphi_\infty$}}
\put(25,170){\makebox(2,-1){\small$\sqrt r$\footnotesize$W$, \footnotesize$H$}}
\put(253,170){\makebox(2,-1){\small$\varphi\!-\!\varphi_\infty$}}
\end{picture}
 \caption[0ba2]{\label{0ba2}
  The fundamental and the first six excited monopole solutions for
  $\alpha=0.2$ and $\beta=0$}
\end{figure}

 If one shifts, however $\varphi_\infty$ to 0 then all of the excited
 monopoles merge with the finite energy abelian solution for
 $\alpha\to\sqrt3/2$.
 The fundamental and the excited monopoles up to six zeros are plotted on
 Fig.~\ref{0ba2}. In order to better display the higher zeros of $W$ we
 plotted $\sqrt r W$. Notice that the newer zeros appear nearer and
nearer to the origin.
 
 The structure of the solutions can be understood from the linearization
 of the field equations (\ref{dymhanzeq})
 around the abelian monopole (\ref{dymhgenrn}). Using the
 \begeq{dymhrnlinanz}
   W=w(r)\;,\quad H=1+{h(r)\over r}\;,\quad
   \varphi=\varphi_{\rm a}(r)+\psi\;,
 \endeq
 variables, from the linearized field equations one finds for
 ${1/M}\ll{\alpha /r}\ll{1/\abs A}$ (where
 $\varphi_{\rm a}\simeq\ln(r/\alpha)/\alpha$) that the solutions are
 well approximated by:
 $\psi\sim e^{\lambda_\psi\tau},\; h\sim e^{\lambda_h\tau},\;
 w\sim e^{\lambda_w\tau}$ where$\tau=\ln r$ and the `frequencies' are
 \begeq{dymhimzmod}
    \lambda_\psi=(1;-2)\;,\quad \lambda_h={1\over2}(1\pm
    \sqrt{1+4\alpha^2\beta^2})\;,\quad
    \lambda_w=-{1\over2}\pm\sqrt{\alpha^2-3/4}\,.
 \endeq
 So we see that in this `middle' region $W$ oscillates with an amplitude
 decaying like $1/\sqrt r$.
 If $M\to\infty$ (i.e. $\alpha E\to1$) this region streches out
 to infinity, while 
 $W$ has more and more zeros when $\alpha<\sqrt3/2$.
 This behaviour is a similar to the one found in the DYM case
 in the interval
 where $\varphi$ grows logarithmically.
 If $\alpha>1$ $W$ 
 we do not expect
 the corresponding solution to exist.
 In the asymptotic region defined by $r\gg\abs{\alpha A\coth(A/M)}$,
 $r\gg\mu^{-1}$
 (where $\mu=\abs{\sinh(A/M)/A}$ and 
 $\varphi_{\rm a}\simeq -{1\over\alpha}\ln\mu$)
  the linear corrections $\beta\not=0$ are characterized
 by $\psi\sim e^{\lambda_\psi\tau},\;w\sim e^{\lambda_w r},\;
 h\sim e^{\lambda_h r}$, with
 \begeq{dymhaszmod}
  \lambda_\psi=(0;-1),\quad\lambda_h=\pm\mu\abs\beta,\quad
  \lambda_w=\pm\mu,
 \endeq
 while $h\sim e^{\lambda_h\tau}$ and $\lambda_h=(0;+1)$ for $\beta=0$.
 
\begin{figure}
 \hbox to\hsize{\hss
  \figbox{0.5\hsize}{70 72 818 556}{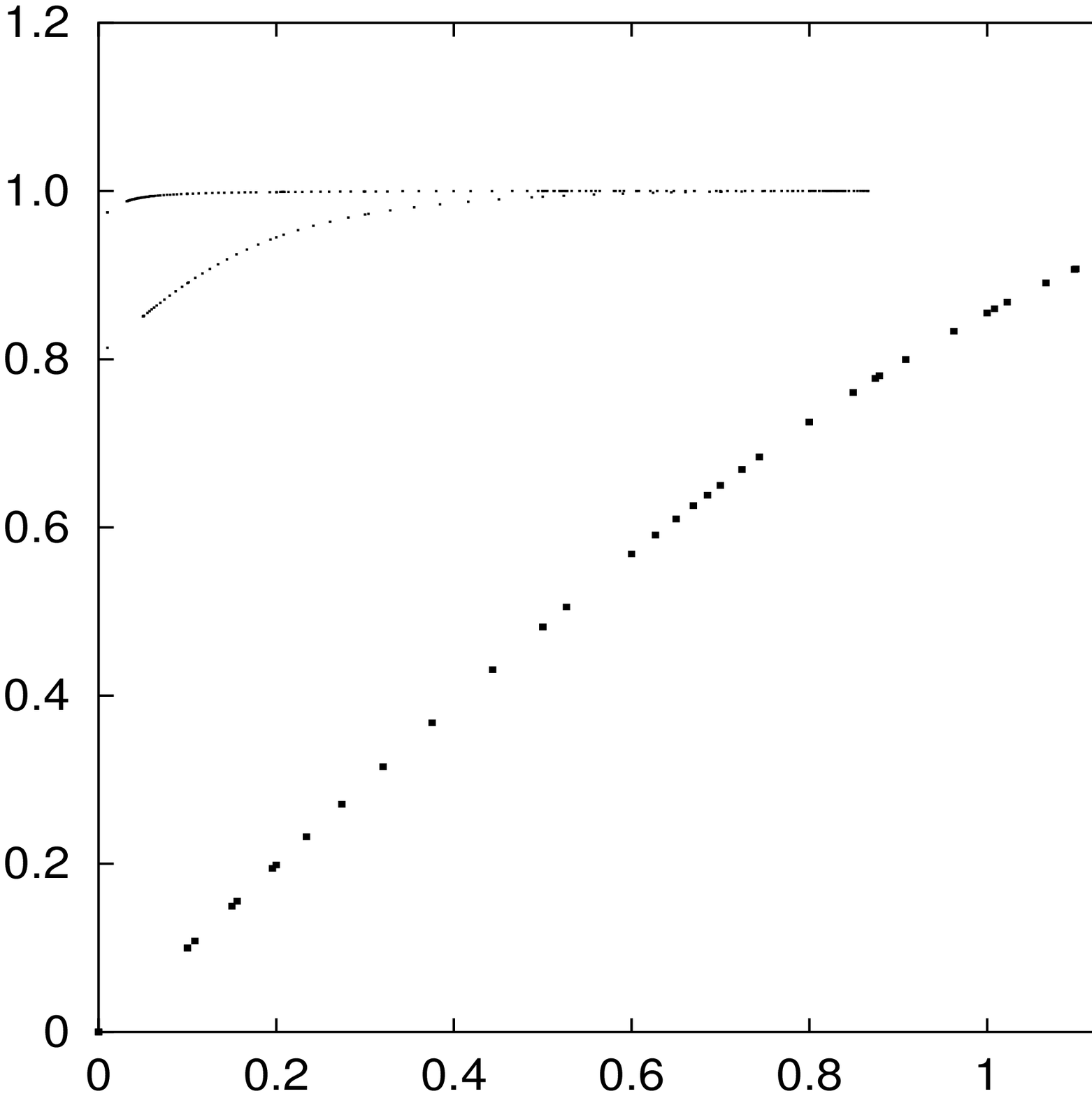}\hss
  \figbox{0.5\hsize}{70 72 818 556}{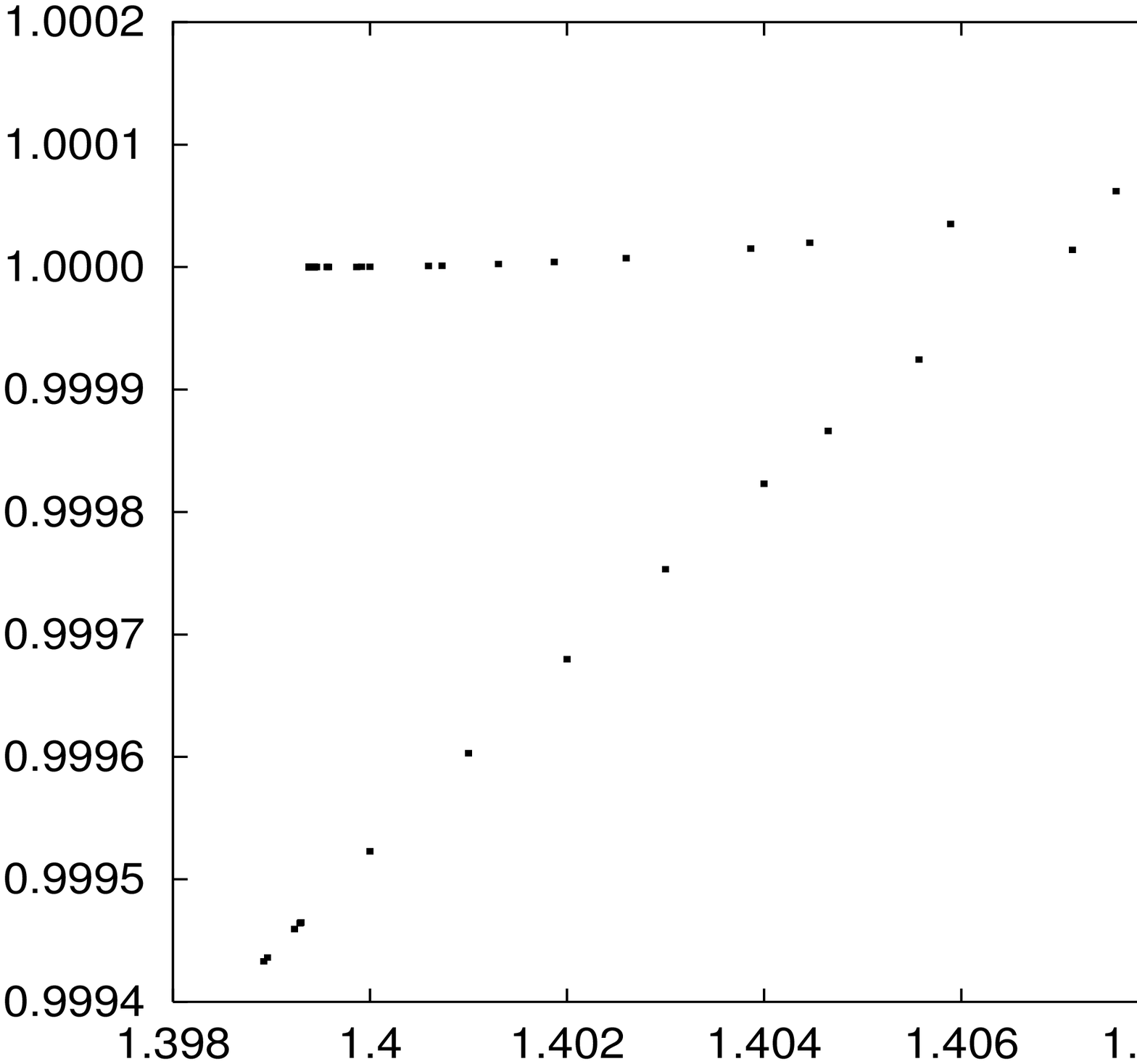}\hss }
\begin{picture}(0,0)(0,0)
\put(205,0){\makebox(0,0){\small$\alpha$}}
\put(440,0){\makebox(0,0){\small$\alpha$}}
\put(25,165){\makebox(0,0){\small$\alpha$\footnotesize$E$}}
\put(250,165){\makebox(0,0){\small$\alpha$\footnotesize$E$}}
\end{picture}
 \caption[0bea0]{\label{0bea0}
  $\alpha$ dependence of the energy of the fundamental, the first, and
  the second excited monopole solutions, and its
  detailed view for the fundamental
  monopole near $\alpha_{\rm m}$ for $\beta=0$}
\end{figure}

\begin{figure}
 \hbox to\hsize{\hss
  \figbox{0.5\hsize}{70 72 818 556}{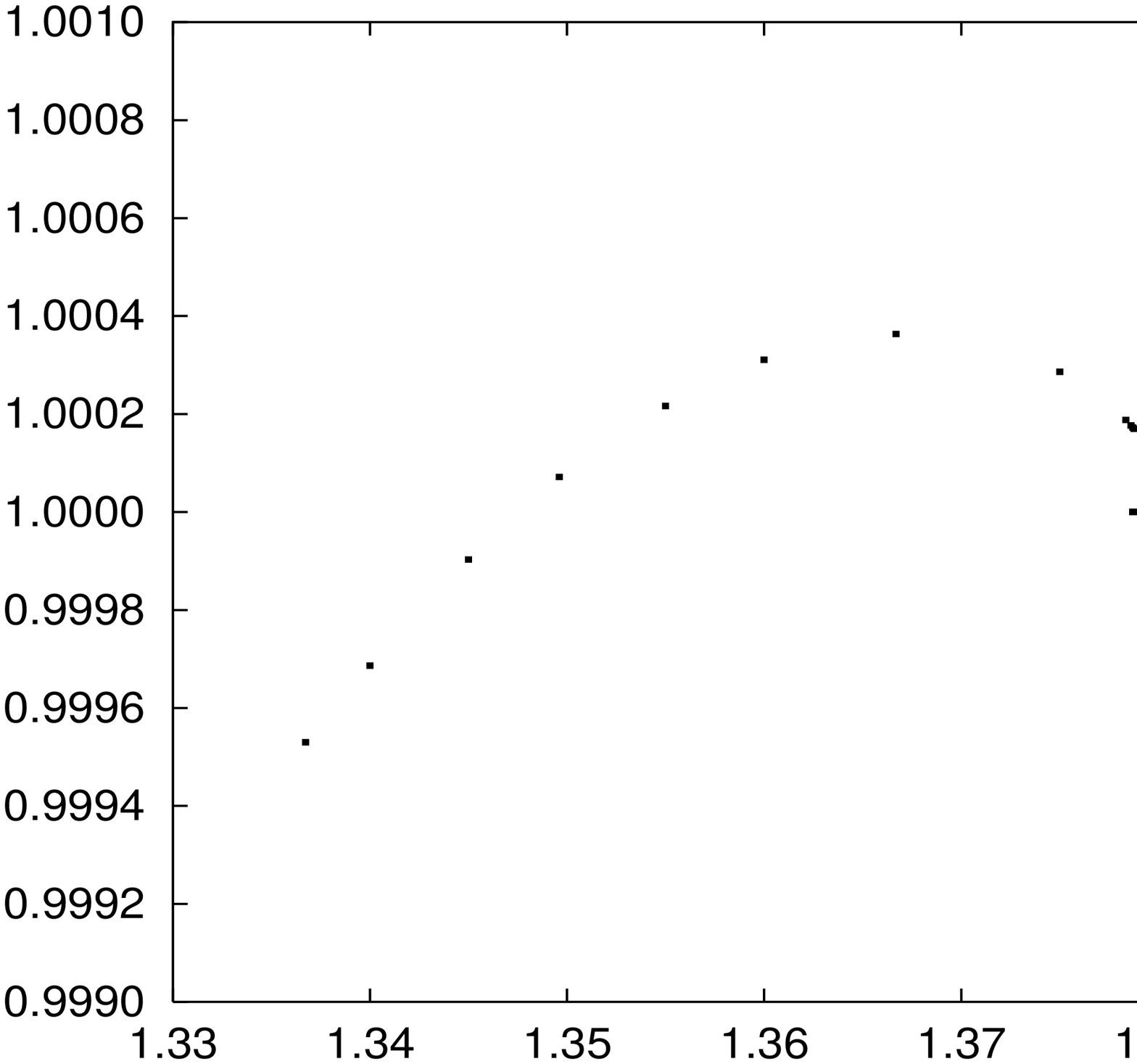}\hss
  \figbox{0.5\hsize}{70 72 818 556}{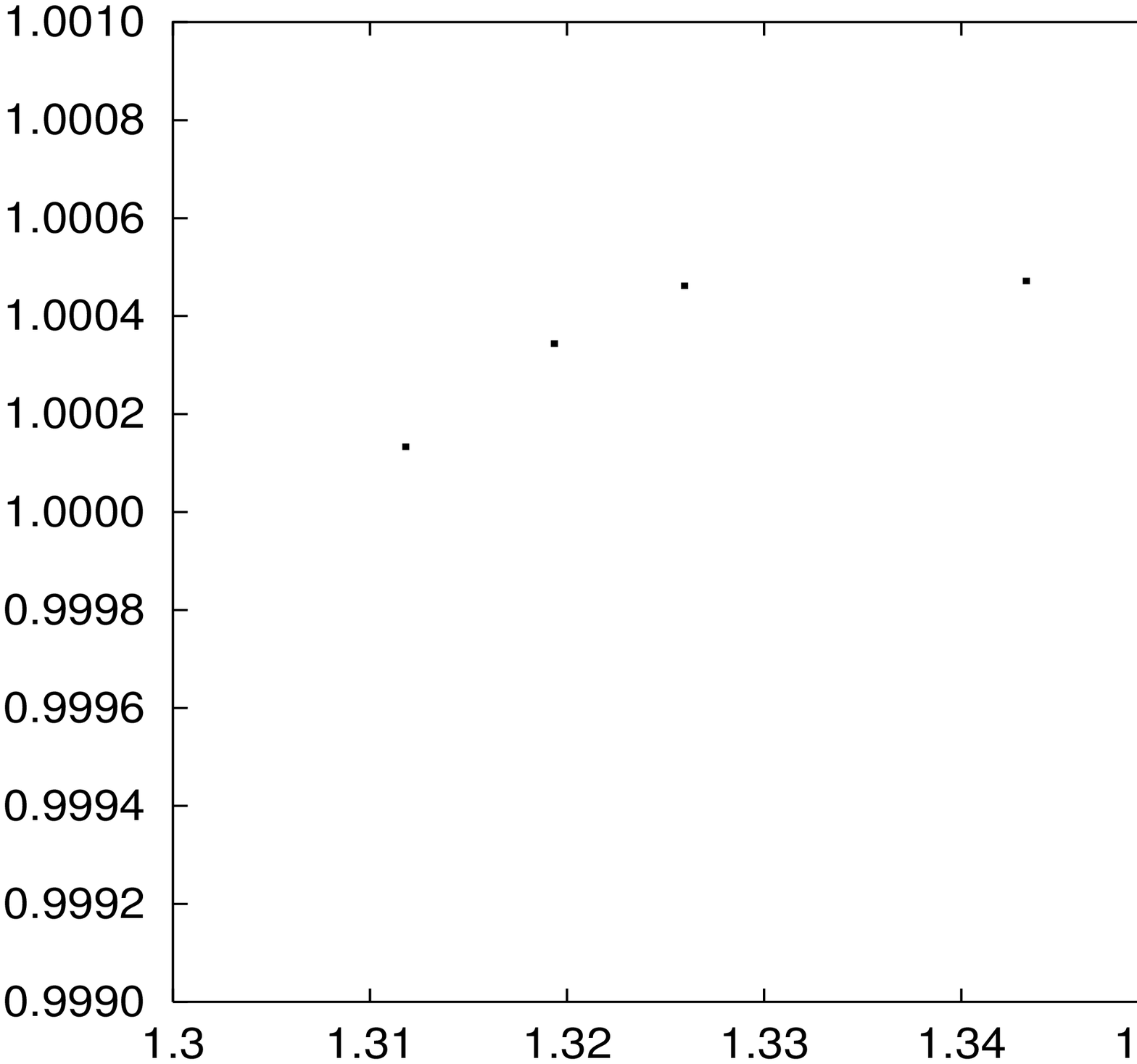}\hss }
\begin{picture}(0,0)(0,0)
\put(205,0){\makebox(0,0){\small$\alpha$}}
\put(440,0){\makebox(0,0){\small$\alpha$}}
\put(25,165){\makebox(0,0){\small$\alpha$\footnotesize$E$}}
\put(250,165){\makebox(0,0){\small$\alpha$\footnotesize$E$}}
\end{picture}
  \caption[b02eaz]{\label{b02eaz}
  Detailed view of the $\alpha$ dependence of the energy of the fundamental
  monopole solution near $\alpha_{\rm m}$ for $\beta^2=0.02$ and for
  $\beta^2=0.043$}
\end{figure}

 The energy of the solutions goes rapidly to $1$ if the number of
 oscillations of $W$ increases (see Table~\ref{T5}).
\begin{table}[htb]
\centering
\caption{}
\vspace{.2truecm}
\label{T5}
\begin{tabular}{|l|l|l|r@{.}l|c|}
\hline
\multicolumn{6}{|c|}{Parameters of the solutions for $\alpha=0.2$ and
  $\beta=0$ in Fig.~\ref{0ba2}}\\
\hline
\multicolumn{1}{|c|}{$n$}&
\multicolumn{1}{|c|}{$a$}&
\multicolumn{1}{|c|}{$b$}&
\multicolumn{2}{|c|}{$\varphi_\infty$}&
\multicolumn{1}{|c|}{$\alpha\cdot E$} \\ \hline\hline
 0 & 0.33297626 & 0.16697893 & 0&1007578  & 0.1988268 \\
 1 & 0.99147112 & 7.52495447  & 10&854544  & 0.9448188 \\
 2 & 2.09466373  & 9.47134195  & 29&998964  & 0.9988219 \\
 3 & 2.14396020  & 9.51995711  & 48&727339  & 0.9999720 \\
 4 & 2.14545164  & 9.52111922  & 67&372640  & 0.9999993 \\
 5 & 2.14549487  & 9.52114737  & 86&014682  & 1.0000000 \\
 6 & 2.14549609  & 9.52114805  & 104&6566   & 1.0000000 \\
\hline
\end{tabular}
\end{table}
We have determined the energy of the solution by fitting the
parameters
$A$, $M$ in Eq.~(\ref{dymhgenrn}) in the asymptotic region
using 
formula (\ref{dymhaszenergscal}). The energy determined
this way contains only exponentially small corrections.
We have also plotted the $\alpha$ dependence
of the energy, $E(\alpha)$, (rescaled to $\varphi_\infty=0$)
on Fig.~\ref{0bea0}
for $\beta=0$.
The similarity of Fig.~\ref{0bea0} to Fig~3 in Ref.~\cite{BFM}
where the masses of {\sl gravitating monopoles} are plotted as a
function of the `gravitational coupling strength',
$M_{\scriptstyle\rm W}/M_{\scriptstyle\rm Planck}$,
is indeed striking.
On Fig.~\ref{b02eaz} $E(\alpha)$ for $\beta^2=0.02$
and 0.043 is shown.
 
 For not too large $\beta$ values ($\beta\leq3$)  
$\alpha E$ of the fundamental monopole becomes
 larger than 1 unlike for the excited ones.
 For $\beta=\infty$ this maximum is $1$.
 
 We make finally some remarks on the stability of the solutions. The
 't Hooft-Polyakov monopoles are stable since they are solutions with minimal
 energy \cite{TFS}. It is natural to expect the fundamental monopoles to
 remain stable in the DYMH case for $\alpha>0$. For sufficiently
 small $\beta$, however,
 where the mass of the fundamental monopole is larger than that of the abelian
 one, the nonabelian solution is expected to become unstable against large
 perturbations. If $\alpha_{\rm c}\not=\alpha_{\rm m}$ there is a bifurcation
 point where the linear stability of the solutions can change. In the EYMH
 case this change of stability has been shown in \cite{H}.
 The excited monopoles are expected to be unstable for all $\alpha$ since
 their energy is significantly larger than that of the fundamental ones.
 This heuristic argument is strengthened by the fact that
 in the $\alpha\to0$ limit their counterparts are known to be
 unstable \cite{BizI,LavI}.


\newpage
 
\begin{thebibliography}{99}
 
\newcommand{\CMP}{\sl Commun.\ Math.\ Phys.\ \bf}
\newcommand{\JMP}{\sl J. Math.\ Phys.\ \bf}
\newcommand{\NPB}{\sl Nucl.\ Phys.\ \bf B\,}
\newcommand{\PLB}{\sl Phys.\ Lett.\ \bf B\,}
\newcommand{\PRD}{\sl Phys.\ Rev.\ \bf D\,}
\newcommand{\PRL}{\sl Phys.\ Rev.\ Lett.\ \bf}
 
\bibitem{Wi}
E.~Witten,
{\PLB 155} (1985) 151
 
\bibitem{BFQ}
C.P.~Burgess, A.~Font and F.~Quevedo,
{\NPB 272} (1986) 661
 
\bibitem{GSW}
M.B.~Green, J.H.~Schwarz and E.~Witten,
Supertring theory, Vol.~2 (C.U.P, Cambridge, 1988)
 
\bibitem{LavI}
G.~Lavrelashvili, and D.~Maison,
{\PLB 295} (1992) 67
 
\bibitem{BizI}
P.~Bizon,
{\PRD 47} (1993) 1656
 
\bibitem{BM}
R.~Bartnik, and J.~McKinnon,
{\PRL 61} (1988) 141
 
\bibitem{THP}
G.~'t Hooft,
{\NPB 279} (1974) 276\\
A.M.~Polyakov,
{\sl JETP Lett.\ \bf 20} (1974) 194
 
\bibitem{BFM}
P.~Breitenlohner, P.~Forg\'acs, and D.~Maison,
{\NPB 383} (1992) 357
 
\bibitem{BFMII}
P.~Breitenlohner, P.~Forg\'acs, and D.~Maison,
{\NPB 442} (1995) 126
 
\bibitem{TFS}
Y.S.~Tyupkin, V.A.~Fateev and A.S.~Shvarts,
{\sl Teor.~Mat.\ Fiz.\ \bf 26} (1976) 397
 
\bibitem{H}
H.~Hollmann,
{\PLB 338} (1994) 181
 
\end{thebibliography}
\end{document}